\documentclass[letterpaper, 10 pt, conference]{ieeeconf}  

\IEEEoverridecommandlockouts    
\newtheorem{defn}{Definition}
\newtheorem{lem}{Lemma}

\newtheorem{prob}{Problem}

\newtheorem{thm}{Theorem}
\newtheorem{assum}{Assumption}

\usepackage{cite}
\usepackage{amsmath}
\usepackage{amstext}
\usepackage{amssymb}
\usepackage{tikz}
\usepackage[mathscr]{euscript}
\usepackage{color}
\usepackage{amsfonts}
\usepackage{amsmath}
\usepackage{arydshln}

\usepackage[final]{pdfpages}
\usepackage{csquotes}
\usepackage{breqn}
\makeatletter
\newcommand*{\rom}[1]{\expandafter\@slowromancap\romannumeral #1@}
\makeatother
\usepackage{graphicx}      

\usepackage{xparse}
\usepackage{mathdots}
\usepackage{epstopdf}

\DeclareMathAlphabet\mathbfcal{OMS}{cmsy}{b}{n}
\usepackage{array}

\usepackage{comment}

\usepackage{mathtools,xparse}

\usepackage{pdfpages}
\usepackage{xcolor}
\usepackage{subcaption}

\let\labelindent\relax
\usepackage{enumitem}

\DeclareMathOperator{\diag}{diag}

\begin{document}
\author{
        Philip.~E.~Par\'e,~Sebin~Gracy,~Henrik~Sandberg
        ~and~Karl~Henrik~Johansson}

\title{ Data-Driven Distributed Mitigation Strategies and Analysis of Mutating Epidemic Processes 
}
\maketitle
\begin{abstract}
In this paper we study a discrete-time SIS (susceptible-infected-susceptible) model, where the infection and healing parameters and the underlying network may change over time. 
We provide conditions for the model to be well-defined and study  its stability. 
For systems with homogeneous 
infection rates over symmetric graphs,
we provide a sufficient condition for global exponential stability (GES) of the healthy state, that is, where the virus is eradicated.
For systems with heterogeneous virus spread over directed graphs, provided that the variation is not too fast, a sufficient condition for GES of the healthy state is established.
Appealing to the first stability result, we present two data-driven mitigation strategies that 
set the healing parameters in a centralized and a distributed manner, respectively, in order to drive the system to the healthy state.
\end{abstract}


\section{Introduction}

In December 2019, a novel coronavirus (COVID-19) was detected in Wuhan, China. This virus quickly spread throughout China, and before long, cases were reported across Asia. In March 2020, the World Health Organization (WHO) officially declared COVID-19 a pandemic \cite{who_pandemic}.  While the economic ramifications of COVID-19 have been significant \cite{coronaecon}, the biggest cause of concern remains the growing number of fatalities worldwide, with over $35,000$ deaths being reported \cite{dong2020interactive} as of March 30, 2020, and more fatalities daily. One of the factors that contributed to COVID-19 becoming a pandemic was the ability of individuals who were asymptomatic, or only slightly sick, to travel easily and great distances while being contagious. 
In this work we propose a model that allows for time-varying graph structure as well as mutating virus parameters. While COVID-19 does not appear to be a susceptible-infected-susceptible (SIS) virus, since infected individuals do not appear to become susceptible once they recover, other viruses and diseases such as the common cold, influenza, gonorrhea, and chlamydia, are of an SIS-type.  
Similar to influenza, it is possible that the novel coronavirus will mutate and therefore become more of an SIS virus. Consequently, we focus on mutating SIS models in this paper.

Network-dependent SIS models with static graphs have been studied extensively in the literature in continuous-time \cite{yorke2,khanafer2016stability,nowzari2016analysis} 
and discrete-time \cite{ahn2013global,wang2003epidemic,peng2010epidemic,pare2018analysis}. These models have been extended to include time-varying graph structure for continuous-time models \cite{BokharaieMTNS10,StarniniMC13,pare2015stability,pare2018epidemic} and discrete-time models \cite{prakash2010virus,bokharaie2010spread,gracy2020asymptotic}. 
The work that most closely relates to the present paper is \cite{gracy2020asymptotic}, wherein discrete-time periodic SIS models have been considered. 
Unlike \cite{gracy2020asymptotic}, here we do not impose any periodicity assumptions. However, for heterogeneous virus spread over directed graphs, we assume that the variations in the topology are not  \emph{too fast}, whereas in \cite{gracy2020asymptotic} no such restrictions are imposed.




With respect to mitigating virus spread, there are a number of works in the literature \cite{wan2007network,wan2008designing,enns2012optimal,xu2014adaptive,PreciadoTCNS14,mai2018distributed,acc_multi,liu2019analysis,gracy2020asymptotic}. 
The main goal of these works is to drive the system to the healthy state, the disease-free equilibrium (DFE), or the origin; note, we use these terms interchangeably throughout the paper. 
Many of the previous techniques devise strategies for adjusting the healing rate in order to eradicate the virus, which can be interpreted as treatment efforts and/or antidote administration. We follow in the same vein and propose a technique that is inspired by \cite{gracy2020asymptotic}; however that technique required knowledge of the infection parameter, which we relax here. The strategies proposed in this paper depend on local network structure, the sampling parameter, and the state of the system (global knowledge for one algorithm and only local for the second), and are therefore driven by the virus spread data.

\subsection*{Paper Contributions}
The overarching goal of the paper is to develop a data-driven control strategy which ensures that a mutating epidemic spreading over a time-varying network is eradicated. Towards this end, we first establish sufficient conditions for global -- i.e., with respect to the physical meaning of the model considered --  exponential stability (GES) of the DFE, and subsequently exploit the conditions to devise control algorithms. 
Consequently, the main contributions of the present paper are as follows:
\begin{enumerate}[label=\roman*)]
 \item We establish sufficient conditions for GES of the healthy state, that is, the origin or where the virus dies out; see Theorems~\ref{thm:Main:Result1} and~\ref{thm:Main:Result2}.
 \item We also provide two data-driven mitigation strategies for adjusting the healing rates, one that requires centralized state information and one distributed technique, that ensure the virus is eradicated asymptotically; see Theorems~\ref{thm:cen_con} and~\ref{thm:dis_con}.
 \end{enumerate}

\subsection*{Paper Outline}
The paper is organized as follows. We conclude this section by listing the notation that is used in the paper. 
The problems being investigated are formally presented in Section~\ref{sec:prob:form}. The main findings of this paper are spread out over Sections~\ref{sec:stab:analysis} and~\ref{sec:control}: Section~\ref{sec:stab:analysis} establishes sufficient conditions for GES of the DFE, whereas Section~\ref{sec:control} presents the control algorithms -- both centralized and distributed -- for ensuring that the disease dies out. The theoretical findings are illustrated via simulations in Section~\ref{sec:sim}. Finally, we summarize our main contributions and shed light on some problems of possible interest for the automatic control and mathematical epidemiology communities in Section~\ref{sec:con}.

\subsection*{Notations}
 Let $\mathbb{R}$ and $\mathbb{Z}_{\geq 0}$ denote the set of real numbers and non-negative integers, respectively. For any positive integer $n$, we have $[n] = \{1,\hdots, n\}$. Given a matrix $A \in \mathbb{R}^{n \times n}$, the spectral radius is $\rho(A)$.
A diagonal matrix is denoted as $\text{diag}(\cdot)$. Given a vector $x \in \mathbb{R}^{n}$, its transpose is denoted as $x^\top$.  The Euclidean norm  is denoted by $\left\|\cdot\right\|$, whereas the infinity norm is indicated by $\left\|\cdot\right\|_{\infty}$. The notation $\{A(k)\}_{a}^{b}$ denotes a sequence of matrices $A(k)$, where $k\in \{a, a+1, \hdots, b-1, b\}$.

\section{Problem Formulation}\label{sec:prob:form}

Consider a time-varying network of $n$ agents, where the set of agents remains the same at all times, while the interconnection among the agents are possibly time-dependent. Each agent has its own \emph{infection} rate, denoted by $\beta_i$, and \emph{healing} rate, denoted by $\delta_i$. 
It is possible that the infection and healing rates are also time-dependent. At time $t \in \mathbb{R}$, let $x_i(t)$ denote the infection level of  agent $i$, and let $a_{ij}(t)$ denote the intensity of interconnection between agents $i$ and $j$. The evolution of the infection level of an agent $i$ can, then, be represented as follows: 
\begin{equation} \label{eq:ct}
\dot{x}_{i}(t) = (1-x_{i}) \beta_{i}(t)\sum\limits_{j=1}^{n}a_{ij}(t)x_{j}- \delta_{i}(t)x_{i}(t).
\end{equation}
For every $t \in \mathbb{R}$, $\beta_i(t)>0$, $\delta_i(t)\geq0$ and $a_{ij}(t)\geq0$. Note that for the SIS model considered above, the state  $x_i(t)$ may correspond to an approximation of the probability of infection of $i^{th}$ agent \cite{van2009virus}, or the infected proportion of group~$i$~\cite{fall2007epidemiological}.

While the spread of an epidemic is a continuous process, the data relating to the epidemic is often collected periodically. For instance, during the recent outbreak of the novel coronavirus, Italy compiles its reports once every $24$~hours. Such a sampling of the system behavior prompts us to look at \textit{discrete-time} 
time-varying 
SIS models. 
The model is obtained by  applying Euler's method \cite{atkinson2008introduction} to \eqref{eq:ct}:
\begin{align}\label{eq:dt}
x_{i}(k+1) = x_{i}(k) + h\bigg( &(1-x_{i}(k)) \beta_{i}(k)\sum\limits_{j=1}^{n}a_{ij}(k)x_{j}(k) \nonumber \\& -\delta_{i}(k)x_{i}(k)\bigg),
\end{align}
where $h>0$ is the sampling parameter.

Observe that, particularized for the time-invariant case, the continuous-time dynamics (as in~\eqref{eq:ct}) is developed by using mean-field approximation of a Markov chain model; see \cite{pare2018analysis,van2009virus}.
Since the discrete-time version of~\eqref{eq:ct},  is obtained by applying Euler discretization, ~\eqref{eq:dt} is an approximation of an approximation \cite[Remark~1]{pare2018analysis}. The approximation accuracy of \eqref{eq:dt} has been addressed, via simulations, in \cite[Section~2.2.2]{pare2018virus}. 

In matrix form, we can rewrite \eqref{eq:dt} as:
\begin{equation} \label{eq:matrixform1}
x(k+1) = x(k) + h ((I-X(k))B(k)A(k)-D(k))x(k)
\end{equation}
where $X(k) = \diag(x(k))$, $B(k) = \diag(\beta_{i}(k))$, and $D(k) = \diag(\delta_{i}(k))$. $A(k) = [a_{ij}(k)]$, for every $i = 1,2, \hdots, n$ and for every $j = 1,2, \hdots, n$.  Let us define $\bar{B}(k) := B(k)A(k)$, with its entries being denoted as $\bar{\beta}_{ij}(k)$. With this notation in place, \eqref{eq:matrixform1} can be rewritten as:
\begin{equation} \label{eq:matrixform}
x(k+1) = x(k) + h ((I-X(k))\bar{B}(k)-D(k))x(k).
\end{equation}
The spread of a virus over a network can be captured using graphs, where the nodes represent the agents, and the edges denote the interaction among the agents. More precisely,  let $G_k = (V, E_k)$ represent this network, where $V = {1,2, \hdots, n}$, and $E_k = \{(x_{i}, x_{j}) \mid \bar{\beta}_{ij}(k) \neq 0\}$. Note that the subscript $k$ indicates that we allow for different edges at each time step $k$, that is, time-varying graphs, which provides a more realistic model.

We define the {\em healthy state} as the state where $x_{i}(k) = 0$ for all $i$, which, from \eqref{eq:matrixform}, implies that,  for all $i$, $x_{i}(k_{1}) =0$ for all $k_{1} > k$. Given that a population is infected with a virus, our main interest lies  in ensuring that each agent $i$, regardless of whether it is initially
healthy or sick,   converges to the healthy state. Towards this end, we adopt a two-pronged approach. First, we seek to establish conditions that ensure GES of  the DFE. 
Second,   we exploit these conditions to develop control strategies that  eradicate the epidemic asymptotically. 
More formally, we can summarize our objectives as follows:
\begin{enumerate}[label=(\roman*)]
    \item \label{q1} For the system with dynamics as given in \eqref{eq:matrixform}, under what conditions is the healthy state GES?

\item \label{q2} Given the knowledge of the conditions that ensure GES of the healthy state,  devise a \emph{data-driven} approach that employs state information to set the healing rates and ensures that epidemic is eradicated asymptotically.

\end{enumerate}

We make the following assumptions. 

\begin{assum}\label{assum:2}
For every $i \in [n]$, $h\delta_{i}(k) \geq 0$ and for all $j \in [n]$, $\bar{\beta}_{ij}(k) \geq 0$, for every $k \in \mathbb{Z}_{\geq 0}$.$\blacksquare$ 
\end{assum}
Assumption~\ref{assum:2} ensures that the healing (resp. infection) rate of each agent is non-negative.
\begin{assum}\label{assum:3}
For every $i \in [n]$, $h\delta_{i}(k) \leq 1$ and $h\sum\limits_{j}\beta_{ij}(k) \leq 1$, for every $k \in \mathbb{Z}_{\geq 0}$.$\blacksquare$ 
\end{assum}
Assumption~\ref{assum:3} is essential for the model to be well-defined.

Note that $x_i$ is an approximation of the probability of agent $i$ being sick, or can be interpreted as the percentage of  subpopulation $i$ that is infected. 
Therefore, for the model to be well-defined we need the following lemma.
\begin{lem} \label{lem:eqm} Assume $x_{i}(0) \in [0,1]$, for all $i \in [n]$. 
For 
\eqref{eq:matrixform}, under the conditions of Assumptions~\ref{assum:2} and~\ref{assum:3}, $x_{i}(k) \in [0,1]$ for all $i \in [n]$ and $k \in \mathbb{Z}_{\geq 0}$.~$\blacksquare$
\end{lem}
\textit{Proof:} The proof is quite similar to that of \cite[Lemma~1]{pare2018analysis}, and is, in the interest of space, omitted.~$\square$\\
Lemma~\ref{lem:eqm} ensures that, with respect to system~\eqref{eq:matrixform}, the set $[0,1]^n$ is positively invariant, i.e., once a trajectory of \eqref{eq:matrixform} enters the set $[0,1]^n$, it never leaves the set. 
Therefore for the rest of the paper we assume that the condition in Lemma~\ref{lem:eqm} is satisfied, that is, $x_{i}(0) \in [0,1]$, for all $i \in [n]$.

\subsection*{Preliminaries}
In this subsection, we will briefly recall some stability notions and results that are essential for understanding the findings in this paper.


Consider a system, described as follows:
\begin{equation}\label{eq:autosys}
x(k+1) = f(k, x(k)),
\end{equation}
where $f: \mathbb{Z}_{\geq 0} \times \mathbb{R}^{n} \rightarrow \mathbb{R}^{n}$ is  locally Lipschitz. 
An equilibrium of~\eqref{eq:autosys}  is  said to be (uniformly) asymptotically stable if it is (uniformly) stable and (uniformly) attractive. Furthermore, an equilibrium of~\eqref{eq:autosys} is endowed with the property of  GAS (resp. globally uniformly asymptotically stable (GUAS))
if, besides being asymptotically stable (resp. uniformly asymptotically stable), the system converges to that equilibrium for any initial state. 

A stronger property is that of GES, which is defined as follows:
\begin{defn}\label{defn:GES}
An equilibrium point of~\eqref{eq:autosys} is GES if there exist positive constants $\alpha$ and $\omega$, with $0 \leq \omega <1$, such that
\begin{equation}
\left\|x(k)\right\| \leq \alpha \left\|x(k_{0})\right\|\omega^{(k-k_{0})} \hspace{1mm} \forall k,k_{0} \geq 0, \forall x_{k_{0}} \in \mathbb{R}^{n}. \nonumber
\end{equation}
\end{defn}

We recall a sufficient condition for GES of an equilibrium of~\eqref{eq:autosys} in the following proposition:
\begin{lem}{\cite[Theorem~28, Section~5.9]{vidyasagar2002nonlinear}}\label{thm:vidyasagar:GES}
Suppose there exists a function $V: \mathbb{Z}_{+} \times \mathbb{R}^{n} \rightarrow \mathbb{R}$, and constants $a,b,c >0$ and $p>1$ such that $a\left\|x\right\|^{p} \leq V(k,x) \leq b\left\|x\right\|^{p}$, $\Delta V(k,x):= V(x(k+1)) - V(x(k)) \leq -c\left\|x\right\|^{p}$, $\forall k \in \mathbb{Z}_{\geq 0}$, and $\forall x \in \mathbb{R}^{n}$, then $x=0$ is an exponentially stable equilibrium of \eqref{eq:autosys}.~$\blacksquare$
\end{lem}
The  initial values are in the domain $[0,1]^n$, since otherwise they lack physical meaning; see Lemma \ref{lem:eqm}. 
Therefore, it follows that the DFE of system~\eqref{eq:matrixform} is  GES if the condition in Definition~\ref{defn:GES} (resp. Lemma~\ref{thm:vidyasagar:GES}) is satisfied for all $x_{k_{0}} \in [0,1]^n$.
\section{Stability Analysis} \label{sec:stab:analysis}
This section addresses Objective~\ref{q1} stated in Section~\ref{sec:prob:form}. That is, we  establish sufficient conditions for GES of the DFE. Towards this end, we first consider undirected graphs with homogeneous virus spread, and subsequently the more general case, which accounts for directed graphs and heterogeneous virus spread.

We begin by defining the following:
\begin{align*}
M(k): &= I-hD(k)+h\bar{B}(k)\nonumber\\
\hat{M}(k) : &= I + h ((I-X(k))\bar{B}(k)-D(k))\nonumber
\end{align*}
Observe that $M(k)$ is the state matrix obtained by linearizing~\eqref{eq:matrixform} around the DFE, and will play a crucial role in our main results. 
\subsection{Homogeneous virus spread and undirected graph}
In this subsection, under the assumption that  the infection rate is the \emph{same} for every agent, and that the underlying graph is \emph{undirected},  we provide a sufficient condition for exponential convergence to the healthy state.
\begin{thm}\label{thm:Main:Result1}
 Let Assumptions~\ref{assum:2} and~\ref{assum:3} hold. 
Suppose that, for all $k\in \mathbb{Z}_{\geq 0}$, $\beta_{i}(k) = \beta(k)$ for every $i \in [n]$, and that, for all $k\in \mathbb{Z}_{\geq 0}$, $A(k)$ is symmetric. If $\sup_{k \in \mathbb{Z}_{\geq 0}} \rho(M(k)) < 1$, then the healthy state of \eqref{eq:matrixform} is GES.~$\blacksquare$
\end{thm}
In words, Theorem~\ref{thm:Main:Result1} says that for a virus with homogeneous infection parameters spreading over undirected graphs, under Assumptions~\ref{assum:2} and~\ref{assum:3}, if each pointwise eigenvalue of $M(k)$ lies strictly within the unit disk, then the healthy state is GES.\\
\textit{Proof:} Consider the following Lyapunov function, $V(k,x) = \frac{1}{2}x^\top x$. 
For $x \neq 0$ and for each $k \in \mathbb{Z}_{\geq 0}$, we obtain the following:
\begin{align}
\Delta V(k,x) &= \frac{1}{2}(x^\top \hat{M}(k)^\top\hat{M}(k)x - x^\top x) \nonumber\\
&=\frac{1}{2}x^\top(M(k)^\top M(k)-I -2h \bar{B}(k)^\top X(k)M(k) \nonumber \\
&\ \ \ \ + h^{2} \bar{B}(k)^\top X(k)X(k)\bar{B}(k))x . \label{eq:b0} 
 \end{align}
By assumption, $\beta_{i} = \beta$ for all $i \in [n]$, and $A(k)$ is symmetric. This implies that $\bar{B}(k)$ is symmetric, and hence $M(k)$ is symmetric. Hence the matrix $M(k)^\top M(k)-I$ is symmetric. Therefore, by applying  the Rayleigh-Ritz Quotient (RRQ) Theorem \cite{horn2012matrix}, we obtain 
\begin{equation}
x(k)^\top(M(k)^\top M(k)-I)x(k) \leq \rho(M(k)^\top M(k)-I) \left\|x\right\|^{2} . \label{eq:b}
\end{equation}
 Since $M(k)$ is symmetric, $\rho(M(k)^\top M(k)) = (\rho(M(k)))^{2}$. By assumption $\sup_{k \in \mathbb{Z}_{\geq 0}}\rho(M(k)) < 1$, which implies, by the definition of the supremum, that for every $k \in \mathbb{Z}_{\geq 0}$, $\rho(M(k)) < 1$ and hence $(\rho(M(k)))^{2} < 1$, thereby implying $\rho(M(k)^\top M(k)) < 1$. 
 
Applying Weyl's inequalities {\cite[Corollary 4.3.15]{horn2012matrix}} to $M(k)^\top M(k)-I$, we obtain $\lambda_{i}(M(k)^\top M(k)-I) \leq \lambda_{i}(M(k)^\top M(k)) -1$, for $i =1,2, \hdots n$. Since, for every $k \in \mathbb{Z}_{\geq 0}$, $\rho(M(k)^\top M(k)) < 1$, it follows that, for each $k \in \mathbb{Z}_{\geq 0}$, $\rho(M(k)^\top M(k)-I) < 0$. Plugging this back into~\eqref{eq:b} yields: $x(k)^\top(M(k)^\top M(k)-I)x(k) <0$ for $x \neq 0$ and for each $k \in \mathbb{Z}_{\geq 0}$. Therefore, from~\eqref{eq:b0}, it follows that 
\begin{align}
\Delta V(k,x) & < h^2x(k)^\top\bar{B}(k)^\top X(k)X(k)\bar{B}(k)x(k) \nonumber \\
\ \ \ \  &\ \ \ \ -2hx(k)^\top\bar{B}(k)^\top X(k)M(k)x(k) \label{eq:b1}\\
\ \ \ \ & = h^2 x(k)^\top\bar{B}(k)^\top X(k)X(k) \bar{B}(k) x(k) \nonumber \\
 \ \ \ \ & \ \ \ \ - 2h^2 x(k)^\top \bar{B}(k)^\top X(k) \bar{B}(k) x(k) \nonumber \\
  \ \ \ \ &\ \ \ \ -2h x(k)^\top \bar{B}(k)^\top X(k)(I -hD(k)) x(k) \nonumber \\
  \ \ \ \ & \leq h^2(x(k)^\top \bar{B}(k)^\top X(k) X(k) \bar{B}(k) x(k) \nonumber  \\
  \ \ \ \ & \ \ \ \ -2 x(k)^\top\bar{B}(k)^\top X(k) \bar{B}(k) x(k)) \label{eq:b2} \\
  \ \ \ \ & \leq -h^2 x(k)^\top \bar{B}(k)^\top X(k)(I-X(k))\bar{B}(k)x(k) \nonumber \\
  \ \ \ \ & \leq 0 \label{eq:b3}
\end{align}
where \eqref{eq:b2} is due to Assumptions~\ref{assum:2} and~\ref{assum:3}, and \eqref{eq:b3} comes from Assumption~\ref{assum:2} and Lemma~\ref{lem:eqm}. Thus, from Lemma~\ref{thm:vidyasagar:GES} the DFE is GES, thereby concluding the proof.~$\square$ 



\noindent
Note that Theorem~\ref{thm:Main:Result1} is the discrete-time counterpart of   \cite[Theorem~1]{pare2018epidemic}.\\

In practice, it is not necessarily the case that each agent in a population has the same infection  rate at each time step, since, for instance, different agents could have different immunity levels. Moreover, the underlying graphs could also be directed. That is, node $i$ being connected to node $j$ does not necessarily imply that the converse is true, or that the edge weightings are equal.
Hence, in the next section we focus on finding conditions  that \emph{also} factor in 
heterogeneous virus spread over directed graphs.

\subsection{Heterogeneous virus spread and directed graphs}
In this section, we establish a sufficient condition for exponential convergence to the healthy state \emph{without} the assumptions of homogeneous infection rates and symmetric interaction weights. 
Towards this end, we make the following assumption: the variation in 
the virus spread parameters and
the topology is \emph{slow}. A practical motivation for imposing such an assumption stems from the observation that one of the accelerating factors in the spread of an epidemic is the mobility of agents across \emph{large distances in a short amount of time}, as shown during the recent COVID-19 pandemic~\cite{li2020substantial}. 
That is, if these slow time-varying measures had been put into place, the pandemic possibly could have been prevented.


%
\begin{thm} \label{thm:Main:Result2}
 Let Assumptions~\ref{assum:2} and~\ref{assum:3} hold. Suppose that
 \begin{enumerate}[label=\roman*)]
     \item  for some $\alpha_1 >0$, $\sup_{k \in \mathbb{Z}_{\geq 0}} \rho(M(k)) \leq  \alpha_1 
     <1$;
     
     \item there exists $L \in \mathbb{R}_{+}$ such that, for all $k \in \mathbb{Z}_{\geq 0}$, $\lvert|M(k)\rvert| \leq L$; and
     \item there exists $\kappa \in \mathbb{R}_{+}$ such that $\sup_{k \in \mathbb{Z}_{\geq 0}}\lvert|M(k+1) - M(k)\rvert| \leq \kappa$.
 \end{enumerate}
 If $\kappa$ is sufficiently small,  then the healthy state of \eqref{eq:matrixform} is GES.~$\blacksquare$
\end{thm} 
In words, Theorem~\ref{thm:Main:Result2} says that under Assumptions~\ref{assum:2} and~\ref{assum:3}, if the  topology of $G_k$ is not changing too quickly,
each pointwise eigenvalue of $M(k)$ strictly lies within the unit disk, and the sequence of matrices $\{M(k)\}_{0}^{\infty}$ is bounded, then the healthy state is GES.

The proof of Theorem~\ref{thm:Main:Result2}
follows almost immediately from that of  the linear work in   \cite{desoer1970slowly} and also, to a lesser extent, that of \cite[Theorem~24.8]{rugh1996linear}). 

\textit{Proof:} See Appendix.~$\square$
Theorem~\ref{thm:Main:Result2} is the discrete-time counterpart of \cite[Theorem~2]{pare2018epidemic}, and differs from \cite[Theorem~1]{gracy2020asymptotic} since no periodicity assumptions are imposed.
%

 
 
\section{Data-Driven Distributed Mitigation}\label{sec:control}

In this section we address Objective~\ref{q2} stated in Section~\ref{sec:prob:form}
by proposing several 
control algorithms that eradicate the virus, using both centralized and distributed techniques. Notice that the eradication strategies developed in this section only account for homogeneous spread over undirected graphs. 
For the control inputs, we set the healing rates locally, $\hat{\delta}_{i}(k)$. This approach can be interpreted as administering antidote and/or other treatment methods.  
The control inputs are designed, for all  $k \in \mathbb{Z}_{0}$,   $i \in [n]$, as
\begin{equation} \label{eq:control}
\hat{\delta}_{i}(k) =  
\max\left\{\hat{\delta}_{i}(k-1),\psi_i(k)\right\},
  \end{equation}
  where 
  \begin{equation}\label{eq:psi}
      \psi_i(k) = \min\left\{\gamma_i(k)\sum\limits_{j=1}^{n} a_{ij}(k) + \eta_{i},1/h\right\}
  \end{equation}
and $\gamma_i(k)$ is 
a data-driven parameter, because it is updated using state information, and 
 $\eta_{i} >0$, for each $i \in [n]$. 
   
The problem becomes the following:
\begin{prob}\label{prob:control}
Replace $\delta_i(k)$ in \eqref{eq:dt} with $\hat{\delta}_{i}(k,\gamma_i(k),a_{ij}(k))$ 
defined in \eqref{eq:control}. 
Find $f$, where $\gamma_i (k) = f(\gamma_i(k-1),x(k))$, that drives the system to the healthy state. 
\end{prob}

From here on, we use  $\hat{\delta}_{i}(k)$ as shorthand notation for $\hat{\delta}_{i}(k,\gamma_i(k),a_{ij}(k))$.

The first proposed solution to Problem \ref{prob:control} is a centralized scheme that updates $\gamma(k) = \gamma_i(k)$, for all $i\in [n]$, as 
\begin{equation}
    \gamma(k+1) = \gamma(k) + \sum_{i=1}^n x_i(k),
    \label{eq:betahat}
\end{equation}
with $\gamma(0) = 0$. This algorithm, combined with \eqref{eq:control}, implies that a central operator knows the sickness level of every node in the network and broadcasts 
$\gamma(k)$
to each node at every time step. Each node then incorporates that into their healing efforts. 
\begin{thm}\label{thm:cen_con}
Consider 
the system in 
\eqref{eq:dt} and assume that,
for all $i \in [n]$, $k \in \mathbb{Z}_{\geq 0}$,
\begin{enumerate}[label=\roman*)]
\item 
$h\beta(k) \sum_{j}a_{ij}(k) < 1$, 
\item $\beta_{i}(k) = \beta(k)>0$, 
and
\item  $A(k)$ is symmetric and $a_{ii}(k)>0$.
\end{enumerate}
Then the algorithm in \eqref{eq:control}-\eqref{eq:psi}, 
with $\psi_i(k) = \psi(k)$ for all $i \in [n]$, and 
\eqref{eq:betahat} 
guarantees GAS of the healthy state.~$\blacksquare$
\end{thm}
\textit{Proof:} If there exists $i \in [n]$ such that $x_{i}(0) > 0$, by \eqref{eq:betahat}, $\gamma(k)$ will increase and, as a consequence of \eqref{eq:control}-\eqref{eq:psi} and the assumption that $a_{ii}(k)>0$ for all $i \in [n]$, $k \in \mathbb{Z}_{\geq 0}$,
 $\hat{\delta}_{i}(k)$ will increase, 
 for each $i \in [n]$, unless $ \psi(k) = 1/h$. 
Hence, either the virus dies out asymptotically,  or by \eqref{eq:control} and \eqref{eq:betahat}, since $\eta_i >0$ for each $i \in [n]$, each $\hat{\delta}_{i}(k)$ will 
equal $1/h$.
Thus, at some finite time $T\geq 0$, $M(T) = hBA(T)$. Therefore, for all $k\in \mathbb{Z}_{\geq T}$, 
\begin{align}
\rho(M(k)) &= \rho(hBA(k)) \nonumber\\
    & \leq \|hBA(k)\|_{\infty} \label{eq:inf}\\
    & < 1, \label{eq:assumOnBetaA}
\end{align}
where \eqref{eq:inf} holds by \cite[Theorem 5.6.9]{horn2012matrix} and 
\eqref{eq:assumOnBetaA} holds by the definition of the infinity norm and the assumption that $h\beta \sum \limits_{j}a_{ij}(k) < 1$, for every $k \in \mathbb{Z}_{\geq 0}$.
Thus by Theorem \ref{thm:Main:Result1}, 
the healthy state is GES for $k\in \mathbb{Z}_{\geq T}$.
Therefore, since $T$ is finite,  the healthy state is GAS.~$\square$\\
Note that the proof does not tell us anything about the stability of the system for $k<T$. In fact, it is very probable that there are many $k<T$  where $\rho(M(k))>1$. 
That is to say, in the time it takes to reach $T$, anything can happen, but after time $T$ we have exponential stability. Therefore, exponential stability coupled with asymptotic time results in asymptotic stability. 

Next, we
present another solution to Problem \ref{prob:control} that updates $\gamma_i(k)$ in a local, distributed fashion:
\begin{equation}
    \gamma_i(k+1) = \gamma_i(k) + x_i(k)  + \sum_{j \in \mathcal{N}_i(k)} x_j(k), 
    \label{eq:betaihat}
\end{equation}
where $\gamma_i(0) = 0$ and $\mathcal{N}_{i}(k)$ denotes the set of neighbors of node $i$ at time $k$, that is, $\mathcal{N}_{i}(k) =\{j \mid a_{ij}(k) \neq 0\}$. 
This algorithm, combined with \eqref{eq:control}, implies that each node only includes information of their own sickness level and that of their neighbors, in their healing efforts. In practice, these are people that do not take outbreaks seriously until they become sick or someone they know becomes sick.
\begin{thm}\label{thm:dis_con}
Consider the system in 
\eqref{eq:dt} and assume that
\begin{enumerate}[label=\roman*)]
\item $0 
\leq h\beta(k) \sum_{j}a_{ij}(k) < 1$ for all $i \in [n]$, 
\item  $\beta_{i}(k) = \beta(k)$ for every $i \in [n]$,
and 
\item $A(k)$ is symmetric, irreducible 
for every $k \in \mathbb{Z}_{\geq 0}$.
\end{enumerate}
Then the algorithm in \eqref{eq:control}-\eqref{eq:psi} and \eqref{eq:betaihat} guarantees GAS of the healthy state.~$\blacksquare$
\end{thm}
\textit{Proof:} 
Consider an arbitrary node $i$. There are three possibilities: 1) node $i$ and none of its neighbors ever become infected 2) node $i$ or one of its neighbors is infected initially, or 3) one of node $i$'s neighbors becomes infected at some time $T_1>0$.
Note that the irreducibility assumption on every $A(k)$ ensures that the underlying graphs, $G_k$, are strongly connected for all $k\in \mathbb{Z}_{\geq 0}$. Therefore, in the first case, the virus has died out. 
In the second and third cases, for each $i$, at some point in time $T_2\geq 0$,
an infected agent will be in the neighborhood of agent $i$ and therefore $\gamma_i(k)$ will increase and, as a consequence of \eqref{eq:control}-\eqref{eq:psi} and the irreducibility assumption of $A(k)$ for all $k \in \mathbb{Z}_{\geq 0}$,
 $\hat{\delta}_{i}(k)$ will increase, for each $i \in [n]$,
  unless $ \psi(k) = 1/h$. 
Hence, since the virus does not die out asymptotically,  by \eqref{eq:control}-\eqref{eq:psi} and \eqref{eq:betaihat}, since $\eta_i >0$ for each $i \in [n]$,  $\hat{\delta}_{i}$ will 
equal $1/h$, for each $i\in [n]$ at some finite time $T\geq 0$.
The rest of the proof is analogous to that of Theorem \ref{thm:cen_con}. 
~$\square$

 Note that the main data-driven aspect of the approaches appears in the update of  $\gamma_i(k)$ (and $\gamma(k)$). However, the control input is also determined by the time-varying graphs. Therefore, the algorithm is shifting resources, as needed, to the most congested areas, captured by the  $a_{ij}(k)$ term in \eqref{eq:control}, as well as to the more sickly areas as captured by the local updates in \eqref{eq:betaihat}.

\section{Simulations}\label{sec:sim}

\begin{figure}
    \centering
    \includegraphics[width=\columnwidth]{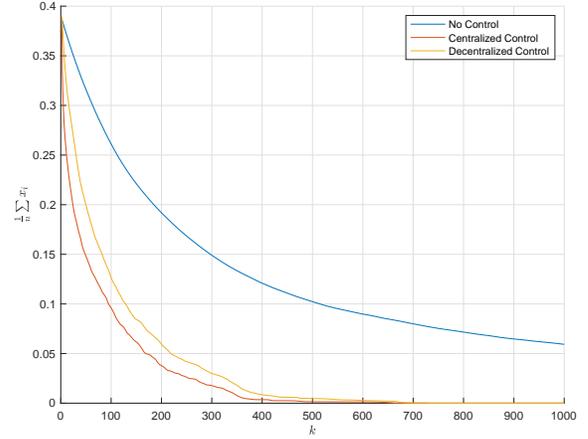}
    \caption{Average infection level of the virus over time}
    \label{fig:time}
\end{figure}

In order to illustrate the results, we simulate a virus spreading over a time-varying graph. The positions of the agents are determined by 
piece-wise constant drifts and the graph structure is determined by the relative 
locations,
similar to that used in \cite{pare2018epidemic}  except with a discrete time framework. 
Namely, this approach confines the agents to a square of edge length $l$ in  $\mathbb{R}^2$, centered at some point $z_c$. So the position of each agent $z_i(k)$ of each agent is updated using following dynamics: 
\begin{equation}\label{eq:phi}
{z}_i(k+1) = {z}_i(k) + \phi_i(k),
\end{equation}
where $\phi_i(k)$ is updated as
\begin{equation}\label{eq:phipiece}
\phi_{ij}(k+1) =\begin{cases}
    -\phi_{ij}(k), & \text{if } z_{ij}(k)=z_{c_j}+l/2 \\
    & \  \text{ or } z_{ij}(k)=z_{c_j}-l/2 \\
   \ \ \phi_{ij}(k) ,             & \text{otherwise},
\end{cases}
\end{equation}
for each dimension $j \in [2]$. That is, if an agent hits a boundary, the velocity of the agent in the dimension corresponding to that boundary flips sign. 
The initial positions are chosen uniformly at random inside the square, each initial $\phi_i$ is chosen uniformly at random, and the initially infected agents are completely infected ($x_i=1$) and are chosen uniformally at random. 
We set $n=1000$ and $\beta = 1$ for all $i\in[n]$. For the simulation without control, we set each $\delta_i$ by sampling a uniform distribution between zero and one. 

\begin{figure}
    \centering
    \includegraphics[width=\columnwidth]{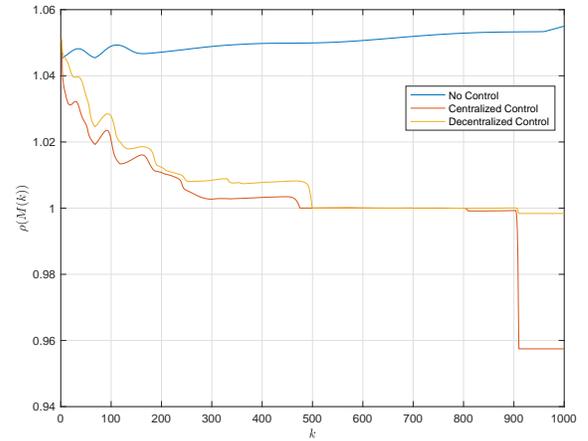}
    \caption{$\rho(M(k))$ of the virus over time}
    \label{fig:spect}
\end{figure}

Using the aforementioned simulation parameters, we simulate three cases: 1) no control, 2) centralized control, as in \eqref{eq:betahat}, and 3) distributed control, using \eqref{eq:betaihat}. The average infection level for each case and $\rho(M(k))$ are plotted in Figures~\ref{fig:time} and \ref{fig:spect}, respectively. Consistent with Theorems \ref{thm:cen_con} and \ref{thm:dis_con}, the control techniques eradicate the virus. As expected, the centralized technique outperforms the distributed technique, however,  not significantly.

\section{Conclusion}\label{sec:con}

In this paper, we considered the problem of eradicating a mutating epidemic in a time-varying network. We first established sufficient conditions for GES of the DFE: simpler conditions for viruses with homogeneous infection rates over symmetric graphs, and more involved conditions for 
heterogeneous virus spread over directed graphs. Subsequently, by leveraging the first stability result, namely Theorem~\ref{thm:Main:Result1}, we devised  two data-driven approaches, one centralised and another distributed, which ensure that the epidemic is eradicated asymptotically. Finally,  we illustrated our theoretical findings via simulations.   

For future work, we would like to extend the ideas herein to SIR (susceptible-infected-recovered) and SEIRS (susceptible-exposed-infected-recovered-susceptible) models so as to possibly better capture the behavior of COVID-19. 
Another possible extension is to derive data-driven mitigation techniques that in addition to eradicating the virus simultaneously \emph{learn} the infection parameter(s) of the system. It would also be interesting to devise  control techniques, similar to those presented in Theorems~\ref{thm:cen_con} and~\ref{thm:dis_con}, that \emph{also} eradicate 
heterogeneous viruses over directed graphs.


\bibliography{ReferencesKTH-Phil}

\section*{Appendix}
\subsection*{Proof of Theorem~\ref{thm:Main:Result2}:}
For each $k 
\in \mathbb{Z}_{\geq 0}$, let $Q(k+1)$ be the solution to the discrete-time Lyapunov equation
\begin{equation} \label{eq:Lyapunov}
    M^\top(k)Q(k+1)M(k) - Q(k+1) = -I_{n} .
\end{equation}
Note that $I_{n}$ is symmetric and positive definite, and, by assumption, $\sup_{k \in \mathbb{Z}_{\geq 0}} \rho(M(k)) <1$. 
Therefore, from \cite[Theorem~23.7]{rugh1996linear}, the solution to~\eqref{eq:Lyapunov} (i.e., $Q(k+1)$) exists, is unique, and is positive definite, for all $k
\in \mathbb{Z}_{\geq 0}$. Furthermore, from proof of \cite[Theorem~24.8]{rugh1996linear}, we have 
\begin{equation} \label{eq:soln:lyapunov}
Q(k+1) = I_{n} + \sum\limits_{j=1}^{\infty}[M^\top(k)]^{j}M^{j}(k). 
\end{equation}
Consider the Lyapunov function $V(k,x) = x^\top(k)Q(k)x(k)$.  
Since $Q(k)$ is positive definite, for all $k
\in \mathbb{Z}_{\geq 0}$, $Q(k)$ is symmetric, by definition. 
Hence, from \eqref{eq:soln:lyapunov},  it is immediate that $I_{n} \leq Q(k)$, for all $k
\in \mathbb{Z}_{\geq 0}$, which further implies $ \lvert|x(k)\rvert|^2 \leq V(k,x)$. 

Next, we focus on finding, independent of $k$, an upper bound on $V(k,x)$. 
Define $\mu = \frac{1-\alpha_1 }{2}$. Therefore, we have $\sup_{k \in \mathbb{Z}_{\geq 0}} \rho(M(k)) \leq 1-2\mu <1$. 
Since by definition, the spectral radius of a matrix is always nonnegative, it follows that $1-\mu >0$. By using Dunford's integral \cite[page 568]{dunford1958linear} with the circle of radius $1-\mu$ as contour, we have, for any positive integer $F$,
\begin{align} \label{eq:dunford}
    M(k)^F &= \frac{1}{2\pi j}\oint\limits_{C}s^F(sI_{n}-M(k))^{-1}ds 
    \nonumber \\
    & \leq \frac{1}{2\pi j}2\pi\lvert s\rvert\max_{\lvert s\rvert =1-\mu}\{s^F(sI_{n}-M(k))^{-1}\}. 
\end{align}
 Taking the norm of both sides of \eqref{eq:dunford} and evaluating at $\lvert s\rvert = 1-\mu$ yields:
\begin{align} 
 \lvert|M(k)^F \rvert| & \leq (1-\mu) \max_{\lvert s\rvert = 1-\mu}(\lvert s \rvert^F) \lvert|\max_{\lvert s\rvert = 1-\mu}(sI_{n}-M(k))^{-1}\rvert| \nonumber \\
 \lvert|M(k)^F \rvert| & \leq (1-\mu)^{F+1}  \lvert|\max_{\lvert s\rvert = 1-\mu}(sI_{n}-M(k))^{-1}\rvert|  \nonumber \\
  & \leq (1-\mu)^{F+1}\max_{\lvert s\rvert = 1-\mu}\lvert|(sI_{n}-M(k))^{-1}\rvert| \nonumber \\
    & \leq (1-\mu)^{F+1}\max_{\lvert s\rvert = 1-\mu}\Big\{\frac{\lvert|(sI_{n}-M(k))\rvert|^{n-1}}{\lvert \text{det}(sI_{n}-M(k))\rvert}\Big\}, \label{eq:norm:kato}
\end{align}
where \eqref{eq:norm:kato} follows from \cite[Lemma 1]{kato1960estimation}.

From \cite[Ex 1,28, page 55]{horn2012matrix}, it is immediate that, for a given $s \in \mathbb{C}$, $\text{det}(sI-M(k)) = (s-\lambda_{j}(M(k)))^n$. Observe that 
\begin{align}
\lvert s-\lambda_{j}(M(k))\rvert &\geq \lvert\lvert s \rvert - \lvert \lambda_{j}(M(k))\rvert \rvert \label{ineq:1}\\
& \geq \lvert \lvert s \rvert - (1-2\mu)\rvert \label{ineq:2} \\
& = \mu, \label{ineq:3}
\end{align}
where ~\eqref{ineq:1} follows from the reverse triangle inequality,
whereas ~\eqref{ineq:2} is due to the following reason. Since, by assumption and the definition of $\mu$, we have $\sup_{k \in \mathbb{Z}_{\geq 0}} \rho(M(k)) \leq 1-2\mu < 1$, it follows from the definition of the supremum that, for all $k
\in \mathbb{Z}_{\geq 0}$, each pointwise eigenvalue of $M(k)$ satisfies $\lvert\lambda_{i}(k)\rvert \leq 1-2\mu$, for all $i\in[n]$. Finally, ~\eqref{ineq:3} is obtained by evaluating~\eqref{ineq:2} at $\lvert s \rvert=1-\mu$, and, hence we have, for $\lvert s \rvert=1-\mu$, $\lvert \text{det}(sI_{n}-M(k))\rvert \geq \mu^n$. 

Since by assumption there also exists an $L$ such that $\lvert|M(k)\rvert| \leq L$, for all $k
\in \mathbb{Z}_{\geq 0}$, it follows that $\lvert|(sI_{n}-M(k))\rvert| \leq (1-\mu + L)$.
Therefore, ~\eqref{eq:norm:kato} can be rewritten as follows:
\begin{align}\label{eq:desoer}
\lvert|M(k)^F \rvert| & \leq \frac{(1-\mu)^{F+1}}{\mu ^n}(1-\mu + L)^{n-1}.
\end{align}
Define $m = \frac{1-\mu}{\mu^n}(1-\mu + L)^{n-1}$ and $p= (1-\mu)$. Thus, \eqref{eq:desoer} can be rewritten as:
\begin{align}\label{eq:desoer:1}
\lvert|M(k)^F \rvert| & \leq mp^F \hspace{4mm} \forall F, \forall k
\in \mathbb{Z}_{\geq 0}.
\end{align}
Taking the norm of both sides of \eqref{eq:soln:lyapunov}, applying the triangle inequality and the submultiplicativity of matrix norms, yields:
\begin{align} 
\lvert|Q(k+1)\rvert| & \leq 1+ \sum_{j=1}^{\infty}m^2 p^{2j} \label{desoer:upper:bound}\\
& \leq \frac{m^2}{1-p^2},\label{desoer:upper:bound2}
\end{align}
where \eqref{desoer:upper:bound2} follows from the fact that \eqref{desoer:upper:bound} is convergent, which can be seen by invoking \eqref{eq:desoer:1} and observing that 
   $p <1$ implies $p^2<1$. 
   Since $Q(k)$ is symmetric for all $k 
   \in \mathbb{Z}_{\geq 0}$, by applying RRQ we have:
   \begin{equation}
   \lambda_{\min}(Q(k))I \leq Q(k) \leq  \lambda_{\max}(Q(k))I, \nonumber
   \end{equation}
   which implies
    \begin{align}
   \lambda_{\min}(Q(k))\lvert|x(k) \rvert|^2 &\leq x(k)^\top Q(k) x(k) \nonumber \\ &\leq  \lambda_{\max}(Q(k))\lvert|x(k) \rvert|^2 \nonumber\\
   & \leq \lvert|Q(k)\rvert|\cdot \lvert|x(k) \rvert|^2 \label{eq:RRQ:1}\\
   & \leq \frac{m^2}{1-p^2} \lvert|x(k) \rvert|^2, \label{eq:RRQ:2}
   \end{align}
   where~\eqref{eq:RRQ:1} follows from  \cite[Theorem 5.6.9]{horn2012matrix}, and ~\eqref{eq:RRQ:2} is due to ~\eqref{desoer:upper:bound2}.
That is, for all $k \in \mathbb{Z}_{\geq 0}$,  $V(k,x) \leq  \frac{m^2}{1-p^2}\lvert|x\rvert|^2$.

For $x \neq 0$ and for all $k \in \mathbb{Z}_{\geq 0}$, we obtain the following:
\begin{align} \label{eqn:deltaV}
\Delta V(k,x) &= x^\top(k)(M^\top(k)Q(k+1)M(k) -Q(k))x(k)
 \nonumber \\
&\ \ \  -2hx^\top(k)\bar{B}^\top(k)X(k)Q(k+1)M(k)x(k)
 \nonumber \\
&\ \ \ + h^{2}x^\top(k) \bar{B}(k)^\top X(k)Q(k+1)X(k)\bar{B}(k))x(k).  
\end{align}
It turns out that $M^\top(k)Q(k+1)M(k) -Q(k)$ is negative definite. To see this, consider the following argument:
Subtracting two successive instances of \eqref{eq:Lyapunov} yields:
\begin{align} \label{eq:diff:1}
M^\top(k)Q(k+1)M(k) - M^\top(k-1)Q(k)M(k-1)
&   \nonumber\\ = Q(k+1) -Q(k)  .
\end{align}
By adding and subtracting $M^\top(k)Q(k)M(k)$ to the  LHS of  \eqref{eq:diff:1}  and rearranging terms, we obtain
\begin{align} \label{eq:diff:2}
&M^\top(k)(Q(k+1)-Q(k))M(k) - (Q(k+1) -Q(k)) 
 =  \nonumber \\ & M^\top(k-1)Q(k)M(k-1) - M^\top(k)Q(k)M(k).
\end{align}
By adding and subtracting $M^\top(k-1)Q(k)M(k)$ to the RHS of  \eqref{eq:diff:2}, we obtain
\begin{align} \label{eq:diff:3}
& M^\top(k)(Q(k+1)-Q(k))M(k) - (Q(k+1) -Q(k)) \nonumber \\ 
 & =  
- ((M^\top(k) - M^\top(k-1))Q(k)M(k)  \nonumber \\ &  \ \ \  \ +  M^\top(k-1)Q(k)(M(k) - M(k-1))) .
\end{align}

Let $R_1 = ((M^\top(k) - M^\top(k-1))Q(k)M(k) + M^\top(k-1)Q(k)(M(k) - M(k-1)))$. Taking the norm of  both sides yields: 
\begin{align} 
\lvert|R_1\rvert| 
 \leq  \lvert| (M^\top(k) - M^\top(k-1))Q(k)M(k) \rvert| + \nonumber \\ \lvert|M^\top(k-1)Q(k)(M(k) - M(k-1))\rvert| \label{tri:ineq}\\
\leq \lvert|(M^\top(k) - M^\top(k-1))\rvert|\cdot \lvert|Q(k)\rvert|\cdot \lvert|M(k)\rvert| \nonumber \\
+ \lvert|M^\top(k-1)\rvert|\cdot\lvert|Q(k)\rvert|\cdot \lvert|M^\top(k) - M^\top(k-1))\rvert|, \label{sub:multi}
\end{align} 
where inequality~\eqref{tri:ineq} is due to the triangle inequality of matrix norms, whereas inequality~\eqref{sub:multi} follows from the submultiplicativity of matrix norms.

Since, for all $k \in \mathbb{Z}_{\geq 0}$, i) there exists 
$\kappa$ such that
$\lvert|M(k+1)-M(k)\rvert| \leq \kappa$, by assumption, and 
ii) $\lvert|Q(k)\rvert| \leq \frac{m^2}{1-p^2}$, by \eqref{desoer:upper:bound2}, 
it follows from \eqref{sub:multi} that $\lvert|R_1\rvert| \leq 2\kappa\frac{m^2}{1-p^2}L$. Notice that \eqref{eq:diff:3} is a Lyapunov equation, whose solution is given by 
\begin{equation} \label{soln:lyapunov:1}
    Q(k+1) -Q(k)  = R_1 + \sum\limits_{j=1}^{\infty}[M(k)^\top]^j R_1 [M(k)]^j . 
    \end{equation} 
    Therefore,     taking the norm of both sides of \eqref{soln:lyapunov:1}, yields 
\begin{align}\label{eq:soln:lyapunov:1:norm:ineq:gp}
\lvert|Q(k+1) -Q(k)\rvert| & \leq \lvert|R_1\rvert|(1+ \sum_{j=1}^{\infty}m^2p^{2j})\\
 & \leq 2\kappa\frac{m^4}{(1-p^2)^2}L.\label{eq:soln:lyapunov:1:norm:ineq}
\end{align}
where inequality~\eqref{eq:soln:lyapunov:1:norm:ineq} is due to \eqref{eq:soln:lyapunov:1:norm:ineq:gp} being convergent. 

Next, pick $\epsilon > 0$ such that $1-\epsilon <1$. Then, from expression~\eqref{eq:soln:lyapunov:1:norm:ineq} it is immediate that if $\kappa \leq \frac{(1-p^2)^2}{2m^{4}L}(1-\epsilon)$, then $\lvert|Q(k+1) -Q(k)\rvert|  \leq 1-\epsilon$. It turns out that  $\lvert|Q(k+1) -Q(k)\rvert|  \leq 1-\epsilon$ implies, for $x \neq 0$ and for all $k \in \mathbb{Z}_{\geq 0}$, \begin{equation}\label{eq:show:neg:def}
x^\top(k)M^\top(k)Q(k+1)M(k) -Q(k)x(k)<0.
\end{equation} 
 To see this, consider the following:
Note that \eqref{eq:Lyapunov} can be rewritten as:
   $ M^\top(k)Q(k+1)M(k) -Q(k) = -I_{n} + Q(k+1) -Q(k)$, for all $k \in \mathbb{Z}_{\geq 0}$. 
Therefore, expression~\eqref{eq:show:neg:def}, for all $k \in \mathbb{Z}_{\geq 0}$, can be written as:
\begin{align} 
&x(k)^\top(-I_{n} + Q(k+1) -Q(k)) x(k) \nonumber \\
& \leq -\lvert|x(k)\rvert^2 + x(k)^\top(Q(k+1) -Q(k))x(k) \nonumber \\
& \leq -\lvert|x(k)\rvert|^2 +
\lambda_{\max}(Q(k+1) - Q(k))\lvert|x(k)\rvert|^2  \label{neg:def:2:1} \\
& \leq -\lvert|x(k)\rvert|^2 + (1-\epsilon)\lvert|x(k)\rvert|^2 \label{neg:def:3} \\
& = -\epsilon \lvert|x(k)\rvert|^2 \nonumber \\
& < 0, \label{neg:def:4}
\end{align}
where 
\eqref{neg:def:2:1} follows from the definition of the induced  norm of $(Q(k+1)-Q(k))^{\frac{1}{2}}$, 
\eqref{neg:def:3} is because a) the norm of a matrix is bounded from below by its spectral radius \cite[Theorem~5.6.9]{horn2012matrix}, and b) $\lvert|Q(k+1) -Q(k)\rvert|  \leq 1-\epsilon$,  and finally 
 \eqref{neg:def:4} follows from the assumption that $\epsilon >0$. 

Since 
 $(M^\top(k)Q(k+1)M(k) -Q(k))$ 
is negative definite, from \eqref{eqn:deltaV}, we obtain, for all $k\in \mathbb{Z}_{\geq 0}$, 
\begin{align} 
\Delta V(k,x) 
& < -2hx^\top(k)\bar{B}^\top(k)X(k)Q(k+1)M(k)x(k)
 \nonumber \\
 & \ \ \ \ + h^{2}x^\top(k) \bar{B}(k)^\top X(k)Q(k+1)X(k)\bar{B}(k))x(k)  \nonumber \\
 & = h^{2}x(k)^\top\bar{B}(k)^\top X(k) Q(k+1) X(k) \bar{B}(k)x(k) \nonumber \\
 &  \ \  \ \ - 2h^{2}x(k)^\top \bar{B}(k)^\top X(k) Q(k+1)\bar{B}(k)x(k) \nonumber \\
 & \ \ \ \ -2hx(k)^\top \bar{B}(k)^\top X(k) Q(k+1) (I-hD(k))x(k) \nonumber \\ 
 & \leq h^{2}(x(k)^\top \bar{B}(k)^\top X(k) Q(k+1) X(k) \bar{B}(k)x(k) \nonumber \\
 & \ \ \ \ - 2x(k)^\top \bar{B}(k)^\top X(k)Q(k+1)\bar{B}(k)x(k)) \label{ineq:key1} \\
 & \leq -h^{2}x(k)^\top \bar{B}(k)^\top X(k)Q(k+1)(I-X(k))\bar{B}(k)x(k) \nonumber \\
 &\leq 0, \label{ineq:key2}
 \end{align}
where \eqref{ineq:key1} is due to Assumptions~\ref{assum:2} and~\ref{assum:3}, and \eqref{ineq:key2} is because of Lemma~\ref{lem:eqm}.~$\square$\\
\end{document}